# Fractal Dimension of Gauge-fixing Defects

M.I. Polikarpov

*ITEP, Moscow, 117259, Russia*

e-mail: polykarp@vxdsyc.desy.de

Ken Yee

*Department of Physics and Astronomy, L.S.U.*
*Baton Rouge, Louisiana   70803-4001, USA*

e-mail: kyee@rouge.phys.lsu.edu

June 14, 1994

## Abstract

The fractal dimension $D_f$ of sites resisting Landau or maximal Abelian(MA) gauge fixing in lattice $SU(3)$ gluodynamics is defined and computed. In Landau gauge such sites clump into $D_f \sim 1$ clusters in the confining phase. In the finite temperature phase their dimensionality drops to $D_f < 1$, that is, clustering seems to dissipate. In contrast, MA gauge resistant sites fail to exhibit a notable tendency to cluster at any temperature.

hep-lat/9402009   14 Jun 94

Gauge fixing is an essential step for an increasing number of lattice QCD applications ranging from the determination of quark wavefunctions to the study of the role of magnetic monopoles in quark confinement [1]. As practitioners know, sites resistant to gauge fixing are routinely encountered in Monte Carlo QCD gauge configurations. In this Note, we describe the space-time distribution of these sites and show, by measuring their fractal dimensionality, that resistant sites are not always randomly distributed. Rather, in Landau gauge they cluster into string-like formations and, further, the fractal dimensionality of these formations is highly temperature sensitive. In contrast, resistant sites in another popular gauge, maximal Abelian gauge, does not clearly exhibit clustering at any temperature.

Landau gauge-fixing in lattice gauge theories is achieved by maximizing the function

$$\mathcal{F}^L \equiv \sum_{x \in \Lambda} F^L(x), \quad F^L(x) \equiv \sum_{\mu=1}^{4} \mathrm{ReTr}(V_x U_{x,\mu} V_{x+\hat{\mu}}^\dagger) \qquad (1)$$

with respect to gauge transformations $V_x$. $U_{x,\mu}$ refers to the link fields. Since $F^L(x)$ at any one site can be maximized exactly by suitably choosing $V_x$, in the simplest algorithms $\mathcal{F}^L$ is maximized iteratively by sweeping through the lattice (perhaps in a checkerboard scheme) and maximizing $F^L(x)$ at each site [2]. However, since nearest neighbor gauge-transformations at $x \pm \hat{\mu}$ ruin the maximization of $F^L(x)$, convergence to a global or even a metastable maximum of $\mathcal{F}^L$ is not guaranteed. Nonetheless, in practice this "relaxation" algorithm does increase $\mathcal{F}^L$ and the gauge configurations do converge to something which at most sites looks like $\partial_\mu A_\mu = 0$.

As previously noted by several authors [2, 3, 4], the relaxation procedure is a limiting case of a Monte Carlo simulation of a quenched Higgs model.

In this formulation, gauge transformations $V_x$ comprise an adjoint $SU(N)$ Higgs field coupled to the gauge fields by the action $S_H \equiv -\kappa \mathcal{F}^L$. $\kappa$ is a real parameter. The Higgs field is quenched because it also interacts (nonlocally) with the gauge fields via the Faddeev-Popov determinant whose *inverse* is

$$\Delta_{FP}^{-1}\big(\kappa; \{U_{x,\mu}\}\big) = \int [dV] \, \exp\{-S_H\}. \tag{2}$$

Landau gauge corresponds to the light Higgs or large $\kappa$ limit; taking $\kappa$ beyond some critically large value $\kappa_c$ forces $\Delta_{FP}$ to be dominated by Higgs configurations which maximize $\mathcal{F}^L$.

At *any* value of $\kappa$, $\Delta_{FP}^{-1}$ can also be viewed as the partition function of an $SU(N)$ spin model with locally variable, pseudo-random couplings $\{\kappa U_{x,\mu}\}$. (Only when $\beta_{QCD} = 0$ are the couplings completely random.) From this viewpoint [5], relaxation corresponds to doing $\kappa > \kappa_c$ Monte Carlo sweeps for the semiclassical ground state spin configurations. Therefore, analogous to defects produced in the instantaneous freezing of liquid iron, defects are expected to be produced during relaxation.

Indeed, a histogram of the $F^L(x)$ distribution after Landau gauge fixing reveals a small but long tail of small-$F^L(x)$ "resistant" sites. As we will show, resistant sites are not homogeneously distributed in space but tend to cluster into spacetime regions, "defects." It is important to get a quantitative handle on defects since they surely affect both action $S_H$ and Faddeev-Popov determinant $\Delta_{FP}$ and, hence, may potentially influence lattice results which depend on gauge fixing.

We must emphasize that, as previously illustrated in Ref [4], some defects are *essential* features of some lattice gauge theories and *not just artifacts of relaxation* which may be potentially suppressed by, for example, an annealing procedure. In compact QED essential defects occur because magnetic

monopoles—which are gauge invariant and physical in this model—populate the gauge configurations. All magnetic monopoles are connected to their antimonopole partners by Dirac strings, which are thusly unavoidable (although gauge variant). Since the Landau gauge vector potential around a Dirac string in the $\hat{z}$ direction is (in continuum cylindrical coordinates) $\vec{A}^{\text{string}} = \hat{\phi}/r_\perp$, the Landau gauge value of $F^L(x) = \sum_{\mu=1}^{4} \cos A_\mu(x)$ must be disrupted near a Dirac string or, equivalently, along a path connecting each monopole-antimonopole pair. As depicted in Figure 1 of Ref. [4], resistant sites cluster around Dirac strings. In $D = 3+1$ dimensional spacetime, Dirac strings sweep out worldsheets which would cause fractal dimension $D_f \sim 2$ Landau gauge defects. Such defects are indicative of the disorder caused by Dirac strings which, for example, gives the photon propagator a nonzero mass pole $M_\gamma$ in the confined phase. $M_\gamma$ is gauge dependent because the location and density of Dirac strings vary with gauge. In the deconfined phase, monopoles and Landau gauge Dirac strings become dilute. Correspondingly, resistance to gauge fixing dissipates and $M_\gamma$ in Landau gauge vanishes.

In this Note we show that the fractal dimension of Landau gauge defects in the confining phase of $SU(3)$ gluodynamics is also nontrivial. Furthermore, in the finite temperature phase we find that these defects tend to dissipate and $D_f$ is dramatically smaller.

For comparison, we also look for defects in maximal Abelian(MA) gauge [6], putatively used to identify nonAbelian magnetic monopoles in 't Hooft's Abelian projection scheme [7]. MA gauge fixing is achieved by maximizing $\mathcal{F}^M \equiv \sum_{x \in \Lambda} F^M(x)$ where

$$F^M(x) \equiv \sum_a \sum_{\mu=1}^{4} \text{Tr}(V_x U_{x,\mu} V_{x+\hat{\mu}}^\dagger \lambda_a V_{x+\hat{\mu}} U_{x,\mu}^\dagger V_x^\dagger \lambda_a) \qquad (3)$$

and $\sum_a$ is the sum over the Cartan generators of the gauge group. Our MA

gauge defects are defined using $F^M(x)$ in complete analogy to Landau gauge defects. As described below, resistant sites do not exhibit pronounce clustering in MA gauge at any temperature. Since both MA and Landau gauge are numerically achieved using the same checkerboard relaxation sweeping scheme, the absence of defects in MA gauge tends to rule out algorithmic artifacts as the cause of clustering in Landau gauge.

A compelling difference between MA and Landau gauge is that, unlike Landau, MA gauge leaves a residual $[U(1)]^{N-1}$ local gauge invariance: the set of all transformations $V_x$ which commute with the Cartan generators in (3). This means $\mathcal{F}^M$ is insensitive to structures which obstruct gauge fixing of the $[U(1)]^{N-1}$ subgroup whereas $\mathcal{F}^L$ is not. Could Landau gauge defects be manifestations of such $[U(1)]^{N-1}$ obstructions? Naively, one might anticipate the following. Consider the world lines of the aforementioned Abelian projection monopoles. Such monopoles, typically identified on the lattice by fixing to MA gauge, are $[U(1)]^{N-1}$ invariant and carry $[U(1)]^{N-1}$ magnetic charge. Every monopole is connected to its antimonopole partner by a $[U(1)]^{N-1}$ *variant* Dirac string. As in compact QED, these Dirac strings would sweep out world sheets with a fractal dimensionality of $D_f \sim 2$. Hence, they might induce $D_f \sim 2$ defects in Landau gauge fixing which do not appear in MA gauge fixing.

Unfortunately, our simulations indicate that the actual situation is not so transparent. As described below, Landau gauge defects have an apparent $D_f$ which is closer to 1 than 2. This discrepancy may rule out Dirac worldsheets as the culprits. The only known $D_f \sim 1$ structures in QCD are the magnetic monopole currents themselves. (Fractal dimensionality of some other QCD features is discussed in [8].) Yet, being $[U(1)]^{N-1}$ *invariant*, there is no obvious reason why they should make defects in Landau but not MA gauge.

Additionally, as monopole identification varies with the gauge of the Abelian projection it is hard to write down a concrete relationship between the MA gauge monopole currents and the Landau gauge defects. As described below, we have performed numerical spacetime correlation studies which indicate that Landau gauge defects are *not* closely correlated with the locations of the MA gauge monopole currents. This negative result does *not* necessarily preclude the possibility that defects and monopoles are manifestations of the same underlying QCD effect. It could simply be that these manifestations shift locations between Landau and MA gauge.

The numerical results to be described are based on the following $\beta_{QCD} = 6.0$ pure gauge $24^3 \times N_t$ $SU(3)$ lattices: eight $N_t = 40$ configurations, nine $N_t = 8$ configurations, five $N_t = 6$ configurations, and eight $N_t = 4$ configurations. $N_t$ is the extent of the Euclidean time direction. At $\beta_{QCD} = 6.0$, the finite temperature transition occurs somewhere in between temperature $\tau \equiv 1/N_t = 1/8$ and $\tau = 1/6$ in lattice units.

Each of the gauge configurations is independently fixed to Landau (and MA) gauge. As such, if the configurations are contaminated with Gribov copies [9] it is likely that the Gribov charge would vary between configurations [5]. Therefore, if Gribov copies have a substantial effect on $D_f$, their influence would show up in the size of our jackknife error bars depicted in Figures 1-3. In fact, our errors are no bigger than errors for comparable gauge invariant quantities which are unaffected by Gribov copies. As it is, we observe no evidence for any anomalous $D_f$ fluctuation which might be due to the presence of Gribov copies in a subset of configurations.

We define fractal dimension $D_f$ of defects operationally as follows. Choose a cutoff value $F_{\min}$ and designate all sites $x$ such that $F^L(x) < F_{\min}$ as resistant. Let $\sum_d$ denote the sum over all resistant sites $d$ and $N_d(r)$ the number of resistant sites inside a $D = 3 + 1$ dimensional hypersphere of radius $r$ centered at resistant site $d$. Then the fraction of resistant sites inside the whole collection of such $3 + 1$ dimensional hyperspheres of radius $r$ is

$$N(r) \equiv \sum_d N_d(r) \Big/ \sum_d 1. \qquad (4)$$

The fractal dimension $D_f$ of defects can be extracted as

$$D_f \equiv \frac{d \log N(r)}{d \log r}. \qquad (5)$$

Expression (5) obviously gives the right average value for the fractal dimensionality if the fractals are sufficiently dilute and $r$ is smaller than the size of the fractals. As $r$ becomes much larger than the mean fractal separation, $D_f$ approaches 4. Figure 1 shows $\frac{d \log \langle N \rangle}{d \log r}$ as a function of $\log(r)$ on the confining $N_t = 40$ lattices for both the Landau and the MA gauge defects. For comparison, the $D_f$ of pseudo-defects comprised of randomly placed, homogeneously distributed sites is also shown. Brackets $\langle \ \rangle$ indicate an average over importance sampling QCD gauge configurations. Only Landau gauge reveals a (nearly) flat nonzero plateau region at small $r$, that is, only Landau gauge exhibits definite clustering. In contrast MA gauge defects and, as required, the pseudo-defects do not exhibit a plateau.

As illustrated in Figure 1, $\log(r/a) = 0.41$ is well inside the $r$-plateau region of $D_f$ for Landau gauge defects. Figure 2 depicts $D_f$ at $\log(r/a) = 0.41$ as a function of $\rho$, the number per unit volume of resistant sites, for Landau gauge defects at four temperatures and the homogeneous distribution. (For Landau gauge $\rho$ is varied by adjusting cutoff $F_{\min}$.) In contrast to the homogeneous distribution where $D_f$ monotonically rises with $\rho$, Landau gauge defects all exhibit a $D_f$ plateau over a range of $\rho$.

As indicated in Figure 2, the same temperature dependence is qualitatively reproduced over a wide range of $\rho$. In Figure 3 we display $D_f$ at $\log(r/a) = 0.41$ and $\rho \sim .005$ for a range of temperatures in Landau gauge and MA gauge. As shown in Figure 3, only $D_f$ in Landau gauge rises dramatically as the temperature is lowered below the critical temperature. This drop correlates the clustering of resistant sites in Landau gauge with the onset of confinement. In contrast, MA gauge defects have $D_f < 1$ at all temperatures.

We look for the aforementioned possible spacetime correlation between the $D_f \sim 1$ Landau gauge defects and the Abelian monopole currents $k_\mu$ in MA gauge as follows. At each dual site $x$ where $k_\mu(x) \neq 0$ we draw a 3-dimensional sphere of radius $r$ in the 3-surface orthogonal to $\hat{\mu}$. Call this collection of spheres $\mathcal{B}(r)$. Counting the fraction of Landau gauge resistant sites inside $\mathcal{B}(r)$, we find that for $\rho \sim .005$ less than 10% of the resistant sites are contained inside $\mathcal{B}(r)$ even if $r = 3$ lattice spacings. Thus, Landau gauge defects are not spatially coincident with MA gauge monopole currents. As previously discussed, this does not necessarily mean they are not manifestations of the same underlying QCD property in two different guises. At this point, we do not know.

In summary, we have shown that the Landau gauge vacuum of lattice QCD has string-like defects. Like the Dirac string-originated defects of compact QED, we suspect these QCD defects indicate the existence of an underlying gauge variant string structure in QCD, be it monopole currents or something else. It will be important to investigate what effects these defects have on QCD correlation functions. Based on the compact QED results of Ref. [4], it is attractive to speculate that such strings may be (partially) responsible for disordering gluon and quark propagators giving rise to the calculated nonzero gluon and quark mass poles in QCD [10].

**Acknowledgments**

MIP acknowledges the LSU Physics Department for its hospitality. It is a pleasure for the authors to thank Dick Haymaker and Lai Him Chan for their encouragement. MIP is partially supported by grant 93-02-3609 of the Russian Foundation for Fundamental Sciences, and by the Japan Society for Promotion of Science Program for Collaboration of Japan-FSU Scientists. KY is supported by U.S. Department of Energy grant DE-FG05-91ER40617. The numerical work was performed at the U.S. National Energy Research Supercomputer Center at Livermore, California.

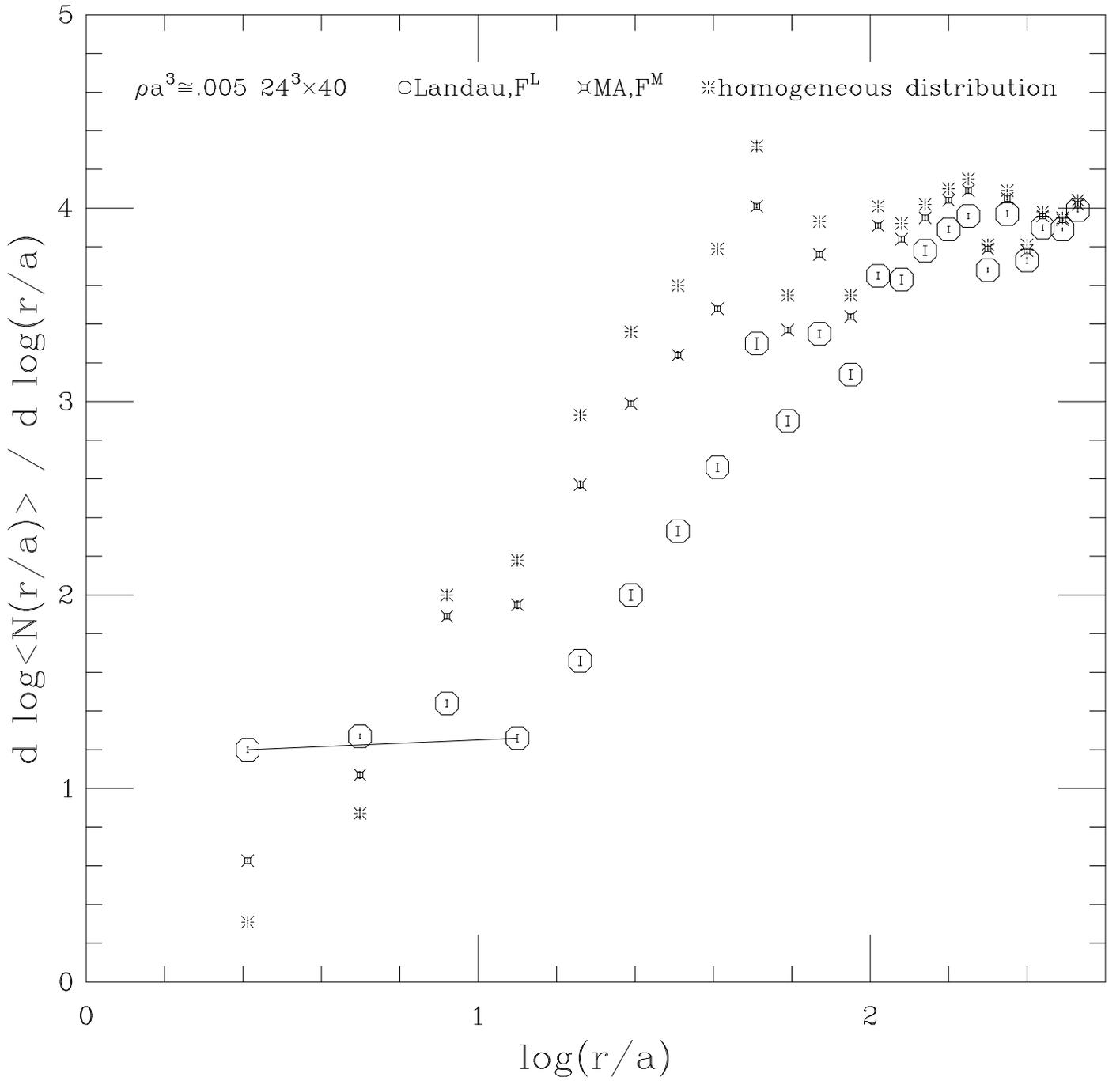

Figure 1: Landau and MA gauge are numerically achieved by maximizing $\mathcal{F}^L$ and $\mathcal{F}^M$ respectively with the same checkerboard relaxation procedure. The Landau gauge curve with resistant sites defined by $F^L$ shows a $D_f \neq 0$ *plateau* over four $r$-slices before beginning to cross over to $D_f = 4$. In contrast, both the MA gauge curve and the homogeneous distribution curve, whose resistant sites are randomly generated with a homogeneous probability, starts crossing over *immediately* from $D_f \sim 0$ at small $r$ to 4 at large $r$ with no plateau region. "a" refers to the lattice spacing at $\beta = 6.0$. Average $\langle\ \rangle$ is over Monte Carlo gauge configurations and all errors are jackknife errors. The guide-to-eye line indicates the plateau region.

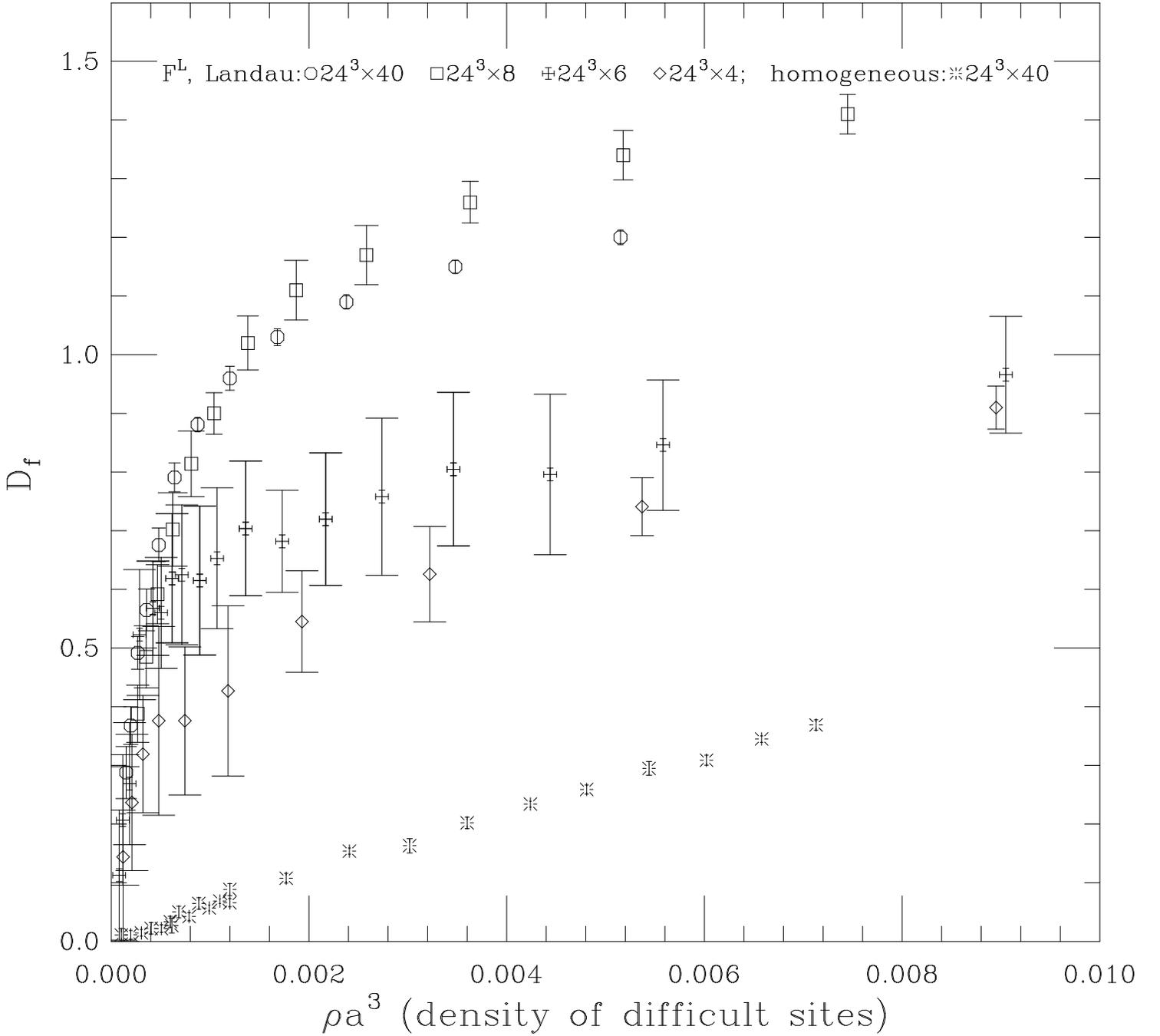

Figure 2: When $\rho$ is too small, $D_f \sim 0$ in all cases. As $\rho$ is tuned upwards, the Landau gauge $D_f$ approaches a nonzero plateau which depends on temperature. These Landau gauge plateau are consistent with clustering. The corresponding MA gauge curves(not depicted) look like the homogenous curve at all temperatures. The jackknife errors on the higher temperature data is larger than the $N_t = 40$ data because the higher temperature lattices contain fewer sites and, hence, worst statistics.

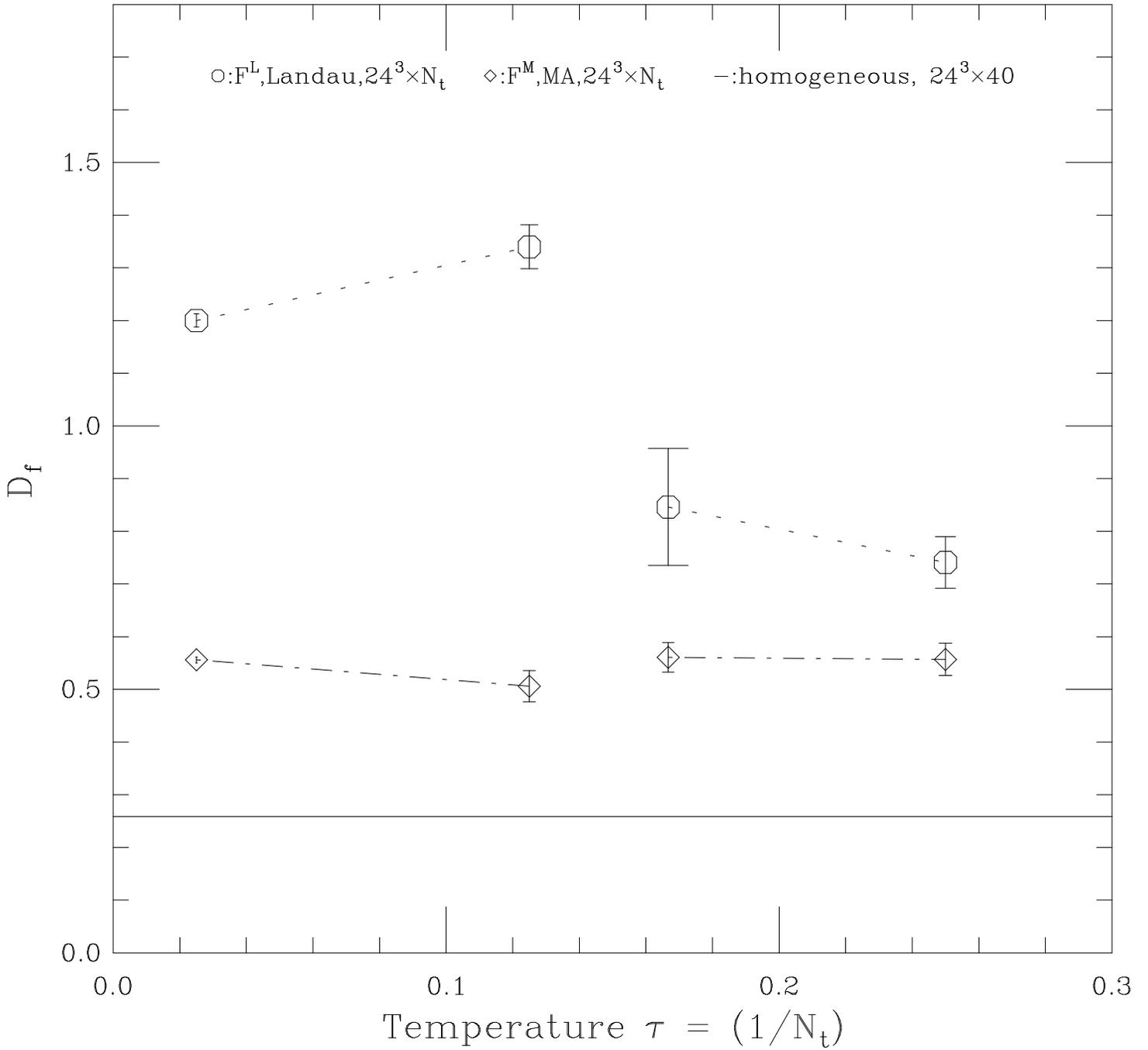

Figure 3: In these $\beta_{QCD} = 6.0$ $SU(3)$ lattices the finite temperature transition occurs somewhere between $\tau = .125$ and $\tau = .16$—the region *not* traversed by the guide-to-eye lines. As depicted, $D_f$ in Landau gauge rises sharply as $\tau$ decreases from the large $\tau$ deconfined phase to the small $\tau$ confined phase. In contrast, $D_f$ in MA gauge is always small. The bold horizontal line is the value of $D_f$ for a homogeneous, random distribution.